\documentclass[12pt,draftcls,onecolumn]{IEEEtran}

\usepackage{amssymb,amsmath,latexsym}
\usepackage{graphicx}
\usepackage{authblk}
\newtheorem{definition}{\textbf{Definition}}

\newtheorem{example}{\textbf{Example}}
\newtheorem{proposition}{\textbf{Proposition}}

\newtheorem{remark}{\textbf{Remark}}
\newtheorem{corollary}{\textbf{Corollary}}

\begin{document}
\title{A Model for Partial Kantian Cooperation\footnote{ This is an ongoing work.\textbf{Any comments or references are very welcome}.}}
\author{Ioannis~ Kordonis, } 
\affil[]{Centralesup\'elec, Avenue de la Boulaie, 35576 Cesson-S\'evign\'e, France\\ email: jkordonis1920@yahoo.com}

\date{\today}
\maketitle

\begin{abstract}
In several game situations, the behavior of the players may depend not only on individual interests, but also on what each player considers as the correct thing to do. This work presents a game theoretic model, aiming to describe game situations in which the players' behavior is affected by ethical considerations. Particularly, we assume that they partially follow, Kant's `Categorical Imperative'. The model is stated for games with a continuum of players. The basic assumption made is that the participants perceive that they belong to virtual (imagined) groups, in which they optimize their actions as if they were bound to follow the same strategy. A partially cooperative equilibrium, called $r$-Kant-Nash equilibrium is then introduced. We then study the relationship of the $r$-Kant-Nash equilibrium with the Nash, (Bentham-) Harsanyi, Rawls difference and Roemer solutions. For the case where the set of possible player types is finite, we prove sufficient conditions for the existence and uniqueness of the $r$-Kant-Nash equilibrium and the equilibrium is characterized in terms of a variational inequality. For the case of continuous types, necessary conditions characterizing the partial Kantian equilibria are derived using a reduction to a set of optimal control problems. Finally, some numerical examples are given.

\end{abstract}

\section{Introduction}

Equilibriua are often inefficient and thus the description of cooperative behaviors has arisen as a major topic in Game Theory. In the context of repeated games there is a lot of work on the imposition of cooperative outcomes, under the name `folk theorems' (see for example \cite{Fud_Tirole}). In the context of Evolutionary Game Theory, the evolution of cooperation is a very important topic as well \cite{axelrod1981evolution},\cite{nowak2006five}.  There is also empirical evidence that people in small societies indeed cooperate, for example when exploiting a common resource \cite{ostrom2015governing}. However, there is a multitude of different cooperative outcomes which can be supported by fully rational players or evolutionary models. Thus,  an important question is `which one of those equilibrium solutions could describe or predict the actual outcomes?'. 

This work studies the behaviour of the players in game situations, in the case where their behavior is affected by ethical considerations. Particularly  we assume that they are partially following Kant's ``categorical imperative'' (\cite{kant1785}): 
\begin{quote}
Act only according to that maxim whereby you can, at the same time, will that it should become a universal law.
\end{quote}
Several issues may arise. First, the players may have different action sets, their actions may have different impact to the others, or they may have different preferences. Hence, a ``maxim'' is interpreted as a strategy of a player (i.e. a mapping from a state or type  to the action set),  and not an action. Second, when a certain player optimizes for the strategy, which she assumes that the others would also follow, it is not reasonable to assume that all the other players having different states or preferences would cooperate to optimize the particular player's cost\footnote{Let us quote a part of a story in which  Woody Allen makes fun of several philosophers. \quote {``No less misguided was Kant, who proposed that we order lunch in such a manner that if everybody ordered the same thing the world would function in a moral way. The problem Kant didn’t foresee is that if everyone orders the same dish there will be squabbling in the kitchen over who gets the last branzino. ``Order like you are ordering for every human being on earth," Kant advises, but what if the man next to you doesn’t eat guacamole? In the end, of course, there are no moral foods-unless we count soft-boiled eggs."}

 From Woody Allen ``THUS ATE ZARATHUSTRA" New Yorker JULY 3, 2006 }. In order to overcome this difficulty, the notion of the veil of ignorance will be used (\cite{harsanyi1980cardinal} under the name equi-partition and \cite{rawls2001justice}). Third, when a veil of ignorance is used the problem of interpersonal comparison of utility arises. However, this is not an issue in a descriptive model, since what is important is how each player perceives the utilities of the others and the players do not need to agree on the scaling of the utilities of the other persons. Finally, the players know that it is not true that all the others will follow their strategy. Hence, it is interesting to study how the players would behave if each one of them assumes that some of the others would  follow her strategy.

In \cite{roemer2015kantian}, \cite{roemerwe}, \cite{roemer2012kantian}, \cite{roemer2013kantian} the notion of the Kantian Equilibrium  is introduced. A set of strategies constitutes a Kantian Equilibrium  if no player has a motivation to change her action assuming that the rest of the players would change their actions accordingly (ex. multiplicatively or additively). It turns out that under weak conditions  the set of Kantian Equilibria coincides with the Pareto frontier. However,  the Kantian Equilibrium is status-quo dependent (and thus conservative) and the players cannot determine their actions, without referring to the actions of the other players. Furthermore, often Pareto frontier contains fundamentally unjust solutions.  It is probably not reasonable to expect that a player who is very much disadvantaged by a solution in the Pareto frontier to be wiling to cooperate with the others, while she has the opportunity to improve her position by changing unilaterally her action. The papers \cite{ghosh2015kant}, \cite{van2015kant}, \cite{grafton2017brave} and \cite{van2017mixed} extended \cite{roemer2015kantian}  in two distinct directions. First, they consider dynamic games and second study game situation with mixed Kantian and Nash players and introduce the notions of (inclusive and exclusive) Kant-Nash equilibria and virtual co-movers equilibrium. Another related line of research is the theory of Belief Distorted Nash Equilibrium \cite{	wiszniewska2016belief}, \cite{wiszniewska2017redefinition}. 

 In this work, we use a model with a continuum of players (see for example \cite{wiszniewska2002static}) and  introduce the notions of $r$-Kant-Nash equilibrium and $r,h$-Kant-Nash equilibrium. we then investigate the  relations of these notions  with several known concepts, including the Nash equilibrium, the Harsanyi solution, the Rawls solution and the  Roemer's Kantian equilibrium. For the finite number of types case, sufficient conditions for the existence and the uniqueness of an $r$-Kant-Nash equilibrium are provided and the $r$-Kant-Nash equilibria are characterized in terms of variational inequalities. For the infinite number of types case, we derive necessary conditions in terms of the solution of appropriate optimal control problems. To illustrate the application of the results, the fishing game example is used. 

 All the sets and functions thereafter are assumed to be measurable.

\section{The Model}

There is a continuum of players, each one of which has an individual type $x_i\in X$ and a social  preference type $\theta_i\in\Theta$. The individual type describes both the preferences of a player and the effects of her actions to the costs of the others (i.e. her position). The social preference type will be explained in detail later on. Denote by $D$ the set of possible individual type-social type pairs i.e., $D=X\times\Theta$ and by $\mathcal{A}$ a sigma algebra on $D$. Let us further denote by $p$ the distribution of the of the pairs $(x,\theta)$ on $D$.

Each player $i$ chooses an action $u_i$ from an action set $U$. The cost function of each player is given by:
\begin{equation}
J_i  =J(u_i,\bar u, x_i),
\label{Cost}
\end{equation}
where $\bar u$ is statistic of the players' actions:
\begin{equation}
\bar u = \int_D g(u(x,\theta),x) dp,
\end{equation}
and by $u(x,\theta)$ we denote the action of a player with the an individual type-social type pair $(x,\theta)$. That is, we focus in symmetric solutions, where all the players with a certain type pair  $(x,\theta)$ use  the same action.

Let us then describe the idea of a virtual (imagined) group:
\begin{enumerate}
\item[(i)]  Each player $i$ assumes that she is associated with a virtual group of players. This group reflects the social considerations of player $i$ and it is constructed according to the social preference type $\theta_i$. This group is described by a sub-probability measure $r(\cdot,x_i,\theta_i)$ on $(D,\mathcal{A})$ i.e., a finite measure with $r(D,x_i,\theta_i)\leq 1$. For every $A\in \mathcal{A}$, the sub-probability measure should satisfy $r(A,x_i,\theta_i)\leq p(A)$. If $r(D,x_i,\theta_i)\neq 0$, let us denote by $ \bar r(\cdot,x_i,\theta_i)$ the probability measure $r(\cdot,x_i,\theta_i)/r(D,x_i,\theta_i)$. If  $r(D,x_i,\theta_i)= 0$ then the virtual group of the player constitutes only of herself. 
% we define $\bar r(\cdot,x_i,\theta_i)=\delta_{x_i,\theta_i}$, where $\delta_{x_i,\theta_i}$ is the Dirac measure.

\item[(ii)] Player $i$ assumes that all the members of the virtual group are bound to use the same strategy $u=\gamma_{x_i,\theta_i}(x)=\gamma(x,x_i,\theta_i)$. That is, Player $i$ assumes that all the players in her group having individual type $x$ will have an action $\gamma(x,x_i,\theta_i)$.

\item[(iii)] The aim of the virtual group is to minimize the mean cost of its members. If $r(D,x_i,\theta_i)> 0$, the virtual group of player $i$ will have a cost:
\begin{equation}
\tilde{J}_i (\gamma,\bar \gamma)= \frac{1}{\beta_{\theta_i}}\ln E\left\{\exp [\beta_{\theta_i} J(\gamma(x',x_i,\theta_i),\bar u_i,x') w_{x,\theta_i}(x')]\right \},
\label{Gr_Opt_Prob}
\end{equation}
where $x'$ is a random variable following the $X$ marginal of $\bar r(\cdot,x_i,\theta_i)$, the factor $w_{\theta_i}$ is a weighting function indicating the relative importance of the several positions in the group and $\beta_{\theta_i}\in[-\infty,\infty]$ is a risk factor. The value of $\bar u$ corresponds to the $g$-mean value of the actions of all the players assuming that the members of the group are using $u=\gamma(x,x_i,\theta_i)$ and the strategy of the players not belonging to the group is given by $\bar \gamma(x,\theta)$. Thus, 
\begin{align}
\bar u = T_{x_i,\theta_i}(\gamma,\bar \gamma) = &\int_D g(\bar \gamma(x,\theta),x) (p(d(x,\theta))-r(d(x,\theta),x_i,\theta_i))+\nonumber \\&~~~~~~~~~~~~~~~~~~~+\int_D g(\gamma(x,x_i,\theta_i),x) r(dx,x_i,\theta_i)
\end{align}
If $r(D,x_i,\theta_i)= 0$, then the cost of the virtual group of player $i$ coincides with the actual cost given by \eqref{Cost}.
If for some $\theta_i$, we have $\beta_{\theta_i}=0$,  then
\begin{equation}
\tilde{J}_i (\gamma,\bar \gamma)=  E\left\{ J(\gamma(x',x_i,\theta_i),\bar u_i,x') w_{x,\theta_i}(x'))\right \}.
\end{equation}
\end{enumerate}

\begin{remark}
\begin{enumerate}
\item[(i)] The virtual groups defined have some points in common with the virtual co-movers model of \cite{ghosh2015kant}. Specifically in the virtual co-movers model each player assumes that if she changes her strategy, a subset of the others would also change theirs accordingly.
\item[(ii)] The definition of the members of the virtual group of each player offers a lot of flexibility. The two extreme cases are the case where $r=0$ and the group of each player consists only of herself and the case where $r(\cdot,x,\theta)=p (\cdot)$. In the intermediate cases the quantity $r(\cdot,x,\theta)$ may relate to some percieved social identity, such as race, class, religion, gender,  ethnicity, ideology, nationality, sexual orientation, culture or language\footnote{ From the identity politics article of Wikipedia}.

\item[(iii)] The virtual groups, the way they are defined, are purely imaginary. Thus, the fact that a player $i$ assumes that an other player $j$ is included in her virtual group does not necessarily imply that the virtual group of player $j$ includes $i$. 

\item[(iv)] It is probably useful to distinguish between the individual position and individual preferences, both contained in $x$. In this case it is possible that within a virtual group the players are allowed to use strategies depending only on their individual position and not their individual preferences. In this case strategies of the form $u=\gamma(y,x_i,\theta_i)$, where $y=h(x,x_i,\theta_i)$ will be considered. 

The players do not necessarily agree on which part of the $x$ variables of the other players correspond to individual positions and which to individual preferences. Let us note that often people tend to overestimate the effect of the character (preferences) of the other persons and underestimate situational factors (position), a phenomenon known as the fundamental attribution error \cite{Aron}.  However, the virtual group of each player belongs exclusively in her perception (or imagination) and thus the function $h$ is defined to depend on $x_i,\theta_i$.

\end{enumerate}
\end{remark}

\begin{definition}
A set of strategies $u=\bar\gamma(x,\theta)$ is an $r$-Kant-Nash equilibrium if for each $(x_i,\theta_i)\in D$ a solution $\gamma(x,x_i,\theta_i)$ of the optimization problem: 
\begin{equation}
\begin{aligned}
& \underset{\gamma(\cdot,x_i,\theta_i)}{\text{minimize ~}}{\tilde{J}_i (\gamma,\bar \gamma)}
\end{aligned}
\end{equation}
 satisfies $\gamma(x_i,x_i,\theta_i)=\bar\gamma(x_i,\theta_i)$.
\end{definition}

A possibly useful alternative is to define the $r$-Kant-Nash equilibrium assuming that the optimization problems in the virtual groups are solved within the class of strategies of the form $u=\gamma(y,x_i,\theta_i)$.

\begin{definition}
A set of strategies $u=\bar\gamma(x,\theta)$ is an $h$,$r$-Kant-Nash equilibrium if for each $(x_i,\theta_i)\in D$ a solution $\gamma(x,x_i,\theta_i)$ of :
\begin{equation}
\begin{aligned}
& \underset{\gamma(\cdot,x_i,\theta_i)}{\text{minimize ~}}{ \frac{1}{\beta_{\theta_i}}\ln E\left\{\exp [\beta_{\theta_i} J(\gamma(h(x',x_i,\theta_i),\theta_i),\bar u_i,h(x',x_i,\theta_i)) w_{\theta_i}(x,x'))]\right \},} \\& \text{subj. to ~~~} \bar u = T_{x_i,\theta_i}(\gamma(h(\cdot,x_i,\theta_i),\bar{\gamma}),  
\end{aligned}
\end{equation}
 within the class of strategies of the form $u=\gamma(y,x_i,\theta_i)$, satisfies $\gamma(x_i,x_i,\theta_i)=\bar\gamma(x_i,\theta_i)$.
\end{definition}

\begin{remark}
Let us comment on the role of the several components of the definitions.
\begin{enumerate}
\item[(i)] The weighting factor $w_{x_i,\theta_i}$ may have two discrete roles. At first, Player $i$ may believe that in her virtual group, some subgroup of players should be favored over the others. A second, and probably more important, role is to resolve the so called \textbf{interpersonal comparison of utility} problem i.e., that fact that the players may not agree on how to scale the utility functions of other players. 
\item[(ii)] The function $h$ has also two possibly distinct roles. At first, player $i$ probably cannot understand or does not know the preference part of the states of the other players belonging to her virtual group. Secondly, it is possible that Player $i$ despite the fact that feels that she belongs to a virtual group involving another player $j$, she does not want to optimize for a certain preference of Player $j$ that she believes as not good or important for the group. 
\end{enumerate}
\end{remark}

\section{Special Cases and Relation to Other Concepts}

The notion of $r$-Kant-Nash equilibrium has several interesting special cases. In the first four cases $\Theta$ is a singleton.
\begin{enumerate}
\item[(i)] The mean field \textbf{Nash equilibrium}. Assuming that $r\equiv 0$ and $\beta_\theta =0$ each player uses her best response to the actions of the other players. Hence, for these values the $r$-Kant-Nash equilibrium coincides with the mean field Nash equilibrium.
\item[(ii)] The \textbf{(Bentham-) Harsanyi} solution. Assume that  $\beta_\theta =0$  and  $r(\cdot,x,\theta) = p(\cdot)$. Then, each player is risk neutral and optimizes for the mean cost (or equivalently the sum of the costs) of all the players. This solution coincides with the solution proposed in \cite{harsanyi1980cardinal}. \sloppy

\item[(iii)] The \textbf{Rawls} solution. Assume that $\beta_\theta =\infty$  and  $r(\cdot,x,\theta) = p(\cdot)$. In this case, all the players minimize the cost function of the worse of participant i.e., they use the minimax rule. This solution coincides with the Rawls difference solution \cite{rawls2001justice}.

\item[(iv)] Assume that $\beta_\theta =-\infty$  and  $r(\cdot,x,\theta) = p(\cdot)$. In this case, all the players minimize the cost function of the \textbf{better off participant}. This optimization procedure makes sense when the better off participant represents the others (for example the best athlete). 

\item[(v)] Efficient cooperation within \textbf{coalitions}. Consider the coalitions $C_1,\dots,C_N \subset D$ and assume that $C_j$, $j=1,\dots,N$ is  a partition of $D$. Further assume that the virtual groups are the same with the coalitions. That is, for $(x_i,\theta_i)\in C_j$ it holds $r(A,x_i,\theta_i)= p(A\cap C_j)/p(C_j)$. Finally assume that $w_{\theta_i}(x,x') = g_{\theta_i}(x')$. Then, within each coalition the players jointly optimize for a weighted sum of their costs, and thus within each coalition there is an efficient cooperation.\sloppy

\item[(iv)] \textbf{Simple Kantian optimization}. The notion of simple Kantian optimization, defined in \cite{roemerwe}, studies cases where each player optimizes for her own cost assuming that all the other players would have exactly the same action with her. This is a limit case of the $h,r$-Kant-Nash equilibrium assuming that $r(\cdot,x,\theta)=p(\cdot)$, the range of the function $h$ is a singleton and $w(x,x') \rightarrow \delta_x(x')$ where $\delta$ stands for the Dirac function. 

\item[(v)] The relation with the altruistic (other regarding) behaviour is illustrated in the following example.

\begin{example}[The Fishing Game]
There is a large number $N$ of fishers each each of which has a cost function:
\begin{equation}
J_i = u_i^2-\left( 1-\frac{1}{N-1}\sum_{j\neq i} u_j   \right)u_i,
\end{equation}
where the first term corresponds to the effort of the fisher $i$ and the second on the revenues. 

The altruistic (other regarding) cost for player $i$ is:
\begin{equation}
\bar J_i = (1-\alpha/2)J_i+(\alpha/2)\sum_{j\neq i} J_j = (1-\alpha/2) (u_i^2-u_i)+\frac{1}{N-1}u_i\sum_{j\neq i}u_j+f(u_{-i}),
\end{equation}
with $\alpha\in[0,1]$. 

The Nash equilibrium of the altruistic game is given by:
\begin{equation}
u_i = \frac{2-\alpha}{6-2\alpha}
\end{equation}

Let us then consider the partial Kantian strategy for the game with a continuum of players. Assume that $\Theta=\{\theta\}$ and $X=\{x\}$ are singletons. Further assume that each player considers as her virtual group a fraction $\alpha$ of the other players. Then, the $r$-Kant-Nash equilibrium  in the form $u=\gamma$ is characterized by:
\begin{equation}
\frac{\partial}{\partial \gamma} J(u,\bar u)=\frac{\partial}{\partial u} J(u,\bar u)+\frac{\partial}{\partial \bar u} J(u,\bar u)\frac{\partial \bar u}{\partial \gamma}=0,
\end{equation}
Hence,
\begin{equation}
2u-1+\bar u +(1-\alpha)u=0.
\end{equation}
Due to symmetry:
\begin{equation}
u = \frac{1}{3+\alpha}.
\end{equation} 
\begin{figure}
\includegraphics[width=\textwidth]{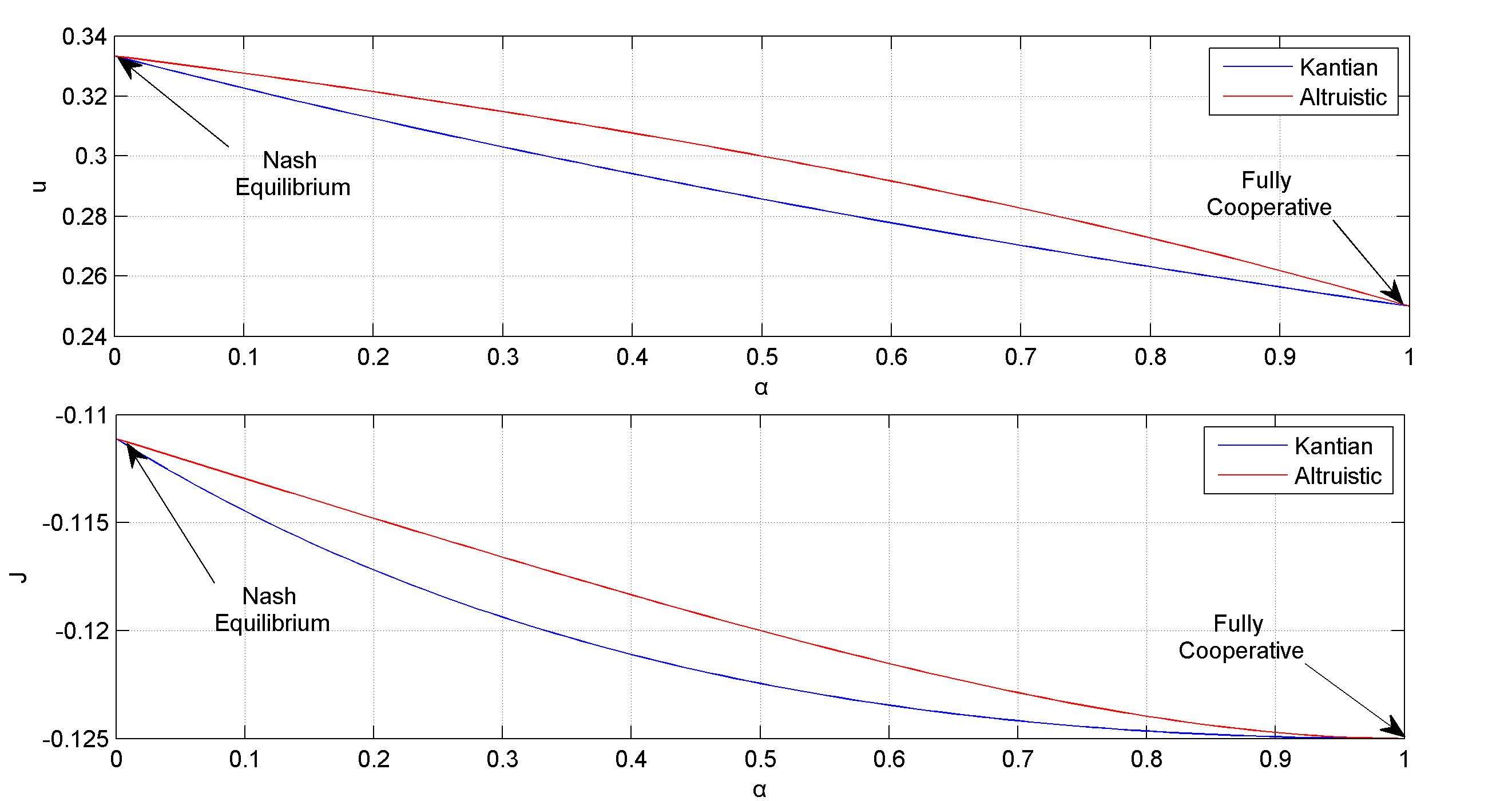}
\caption{Comparison of the actions and the Costs of the players when they use a partially Kantian vs an Altruistic criterion}
\label{Altr_vs_Kant_fig}
\end{figure}
Figure \ref{Altr_vs_Kant_fig} compares the actions and the costs of the players in the cases of an altruistic versus a partially  Kantian behaviour. It turns out that the Kantian cooperation is more effective than the altruism.
\hfill $\square$
\end{example}
\item[(vi)] The relation with \textbf{Roemer's Kantian equilibrium} is illustrated in the following example:
\begin{example} [The Fishing Game: Continued]
Consider again the Fishing Game example.  Assume also that in a previous iteration of the game (status quo) the players used actions $ u_i^{\text{prev}}$. Denote by $x_i$ the state of each player to be the action that she had in the status quo. 

Let the actions of the game to be denoted by $v_i$ and relate to the original Fishing game with $u_i=x_iv_i$. Then, assuming  $\beta_\theta =0$,  $r(\cdot,x,\theta) = p(\cdot)$ and $h(x,x_i,\theta_i)=x_i$ an $h,r$-Kant-Nash equilibrium of the game is a Roemer's Kantian equilibrium.  \hfill $\square$
\end{example} 

This analogy illustrates that the Kantian equilibrium is status quo dependent. Furthermore, the choice of the multiplicative representation of the game is made  such that the outcome belongs to the Paretto frontier. Hence, Kantian equilibrium depends on the ``experience'' in two distinctive ways i.e., the action changing rule and the status quo. Kant in his Groundwork \cite{kant1785} strongly opposes to the use of ``experience'' in order to derive normative rules. However, this fact in no case implies that people are not using their experience to determine how to behave. What it probably means is that in Roemer's work, experience affects too much the behavior of the players, while in this work too little. 
\end{enumerate}

\section{Finite Number of Individual and Social Types}

In this section,  we assume that $U$ is a subset of the $m-$dimensional Eucledian space and that $D$ is finite i.e., there is a finite set of possible individual-social type pairs $(x_1,\theta_1),\dots,(x_N,\theta_N)$. The distribution of players is described by a vector $p=[p_1~\dots~p_N]$. For the case of a finite number of types, we derive sufficient conditions for the  existence of an  $r$-Kant-Nash equilibrium and characterize it in terms of a variational inequality. 
To do so, let us first introduce some notation. Denote by $N'$ the number of different values of $x_k$'s and by $\bar x_{1},\dots,\bar x_{{N'}}$ their values.  Denote also by $\sigma$ the function such that $\sigma(k)=k'$ if $x_k=\bar x_{k'}$.  Consider the virtual group of a player, having type $k$, and assume that this group has strategy $\gamma$. Denote by  $\tilde u_k=[u_{k1},\dots,u_{kN}]$ the (imagined) action vector for the members of the group, where $u_{kk'} = \gamma(x_k',x_k,\theta_k)$. The form of the strategy implies that $u_{kk'}=u_{kk''}$ if $x_{k'}=x_{k''}$. Thus, the vector $\tilde u_k$ may be viewed as a member of the set $U^{N'} $.
 We will use also the notation $r_{kk'}=r(\{(x_k',\theta_k')\},x_k,\theta_k)$.

A vector of actions $u^{\star}=[u^{\star}_1,\dots,u^{\star}_N]$ is an $r$-Kant-Nash equilibrium if there there exists a matrix $u=[u_{kk'}]$ such that,  for every $k$, it holds $ u_k^{\star}=u_{kk}$ and the strategy $\tilde u_k = [ u_{k1}\dots u_{kN} ]$  is optimal for the virtual group of a player with type $k$, under the constraint $u_{kk'}=u_{kk''}$ if $\sigma(k')=\sigma(k'')$. 
The cost of a virtual group with action vector $\tilde u_k$, assuming that the others are playing $u^{\star}$, is given by:
\begin{equation}
\tilde{J}_k(\tilde u_k, u^{\star})= \frac{1}{\beta_{\theta_i}}\ln \sum_{k'=1}^Nr_{kk'}\exp [\beta_{\theta_i} w_{k}(k') \bar J_{k,k'}(\tilde u_{k},u^{\star})],
\label{formula}
\end{equation}
where $J_{ u^{\star},k'}(\tilde u_{k}) $ is the cost that a player belonging to the virtual group having type $k'$ would have if the players of the group were using $ \tilde u_k$ and the rest $u^\star$. Thus, $\bar J_{k,k'}(\tilde u_{k},u^{\star}) $ is given by:
\begin{equation}
\bar J_{k,k'}(\tilde u_{k},u^{\star}) =  J\left(\tilde u_{kk'},\sum_{k''=1}^N [g( u^{\star}_{k''},x_{k''}) (p_{k''}-r_{kk''})+ g(\tilde u_{kk''},x_{k''}) r_{kk''} ]  ,x_{k'}\right).
\label{J_ubar_def}
 \end{equation}

The following proposition adapts some standard results for the existence of a Nash equilibrium (e.g. \cite{debreu1952social}) to the case of $r$-kant-Nash equilibrium. Before stating the proposition let us recall the notions of quasi-convexity and pseudo-convexity \cite{Manga}. A function  $f(u)$ defined on a convex set $U$ is quasi-convex if for any real number $\bar f$ the set $\{u\in U: f(u)\leq \bar f \}$ is convex. The function $f$ is pseudo-convex if it is differentiable and for any pair of points $u_1,u_2\in U$ such that $\nabla f(u_1)^T(u_2-u_1)\geq 0$ it holds $f(u_2)\geq f(u_1)$.

\begin{proposition}
\label{Exist_Prop}
Assume that $U\subset \mathbb{R}^m$ is compact and convex, that $\tilde{J}_{k}$ given by \eqref{formula} is continuous, and that $\tilde{J}_{k} (\cdot,u^{\star})$ is quasi-convex and for every fixed $u^{\star}$ and every $k$. Then, there exists an $r$-Kant-Nash equilibrium.
\end{proposition}
\textit{Proof}: Consider the set $\bar U=U^{N\times N'}$. For each type $k$ consider the correspondence $T_k:\bar U\rightrightarrows U^{N'}$ defined as follows. For a given a $u\in \bar U$, define $u^{\star} =[u_{1\sigma(1)}\dots u_{N\sigma(N)}] $.
Then define:  $$T_k(u) = \pi_{U^{N'}}\left[\underset{\tilde u_k\in Z}{\arg\min} \tilde{J}_k(\tilde u_k, u^{\star})\right],$$ where $\pi_{U^{N'}}$ denotes the projection to ${U^{N'}}$ and  $Z=\{\tilde u\in U^N: \tilde u_{k'}=\tilde u_{k''} \text{ whenever } x_{k'}=x_{k''}\}$. Maximum theorem \cite{berge1963topological} implies that $T_k$ is compact valued upper semi-continuous. Quasi-convexity implies that $T_k$ is convex valued. 
Thus, the correspondence  $T:\bar U\rightrightarrows \bar U$ with  $T:u\mapsto T_1(u)\times\dots\times T_N(u)$ satisfies the conditions of Kakutani fixed point theorem.  Therefore, there exists an $r$-Kant-Nash equilibrium.
\hfill $\square$

A very simple sufficient condition for the existence of a Kant-Nash equilibrium is given in the following corollary.
\begin{corollary}
Assume that $U\subset \mathbb{R}^m$ is compact and convex, and $ J(\cdot,\cdot,x)$ is convex for every fixed $x$, the function $g$ is linear in $u$ and $\beta_{\theta_k}\geq 0$, for all $k$. Then there exists an $r$-Kant-Nash equilibrium.
\end{corollary}
\textit{Proof}: The function $\bar J_{k,k'}(\cdot,u^{\star}) $ defined in \eqref{J_ubar_def} is convex with respect to $\tilde u_{k}$. Indeed in the right hand side of \eqref{J_ubar_def},  the first two arguments of $J $, particularly  
$\sum_{k''=1}^N [g( u^{\star}_{k''},x_{k''}) (p_{k''}-r_{kk''})+ g(u_{kk''},x_{k''}) r_{kk''} ]  $ and $\tilde u_{kk'} $ are affine functions of $\tilde u_k$. Thus, convexity of $J$ implies that $J_{ u^\star k'}$ is convex. Now, the fact that  $\exp(\cdot)$ is increasing and convex implies that the function:
$$\sum_{k'=1}^Nr_{kk'}\exp [\beta_{\theta_i}  J_{u^{\star} ,k'}(\tilde u_{k}))],$$
is convex in $\tilde u_k$ as well. Now, $\beta_{\theta_k}\geq 0$ and the fact that the function $\ln(\cdot)$ is increasing, imply that the quasi-convexity assumption of Proposition \ref{Exist_Prop} is satisfied and the proof of the corollary is complete. \hfill $\square$

If the quasi-convexity assumption is strengthened to a pseudo-convexity, then the $r$-Kant-Nash equilibrium can be characterized by a variational inequality. 

\begin{proposition}
Assume that $U\subset \mathbb{R}^m$ is convex, that $\tilde{J}_{k}$ given by \eqref{formula} is continuous, and that $\tilde{J}_{k} (\cdot,u^{\star})$ is pseudo-convex and for every fixed $u^{\star}$ and every $k$. Consider also the vector function $F:U^{N^2+N}\rightarrow \mathbb{R}^{mN^2+mN}$ given by:
\begin{equation}
F(\tilde u,u^\star)=\left[
\begin{matrix}
\nabla_{\tilde u_1} \tilde J_1(\tilde u_1,u^\star )^T\\
\vdots\\
\nabla_{\tilde u_N} \tilde J_1(\tilde u_N,u^\star )^T\\
u^\star_1-\tilde u_{11}\\
\vdots
\\
u^\star_N-\tilde u_{NN}\\
\end{matrix}
\right]
\end{equation}
 Then:
\begin{itemize}
\item [(i)] There is an $r$-Kant Nash equilibrium if and only if there is a solution to the variational inequality:
\begin{equation}F(\tilde u,u^\star)^T\left  [\begin{matrix}\tilde u'-\tilde u,\\(u^\star)'-u^\star  \end{matrix}\right] \geq 0, \text{~ for all~~}\tilde u'\in U^{N^2},(u^\star)'\in U^{N}\label{VI_Char}
 \end{equation}
\item[(ii)] Assume that $F$ is strictly monotone i.e.: 
\begin{equation}
(F(\tilde u',(u^\star)')-F(\tilde u,u^\star))^T\left  [\begin{matrix}\tilde u'-\tilde u,\\(u^\star)'-u^\star  \end{matrix}\right] >0,
 \end{equation}
for every  pair $(\tilde u,u^\star),(\tilde u',(u^\star)')$. Then, there is at most one $r$-Kant Nash equilibrium.
\end{itemize}

\end{proposition}
\textit{Proof}: (i) Consider a pair $(\tilde u, u^\star)$ satisfying \eqref{VI_Char}. Choosing $\tilde u'=\tilde u$ and $(u^\star_2)'=u^\star_2,\dots,(u^\star_N)'=u^\star_N$ we conclude that $u^\star_1=\tilde u_{11}$. Similarly,  $u^\star_k=\tilde u_{kk}$ for all $k$. Choosing $\tilde u'_i=\tilde u_i$ for $k=1,\dots,k-1,k+1,\dots,N$ we conclude that   $\tilde u_k$ is optimal in \eqref{formula}. Thus, $(\tilde u, u^\star)$  corresponds to an $r$-Kant Nash equilibrium. 
(ii) It is a direct consequence of part (i) and Theorem 2.3.3 of \cite{facchinei2007finite}. \hfill $\square$

\section{ Infinite Number of Types}
\subsection{Reformulation as Optimal Control Problems}

In this section we characterize $r$-Kant-Nash equilibria under the assumption that $\Theta$ is a singleton, $X=[0,{T_f}]$, $\beta=0$,  and that all the measures are absolutely continuous with respect to the Lebesgue measure. 

Assuming that the players that do not belong to $i$'s virtual group follow a strategy $u_j=\bar\gamma(x_j)$, the optimization problem \eqref{Gr_Opt_Prob}
is written as:

\begin{equation}
\underset{{\gamma}}{\text{min.}}  \left\{ \int_0^{T_f} J \left (\gamma(x'),\bar u^{-x_i}+\int_0^{T_f} g(\gamma (z),z)r(z,x_i) dz, x' \right) w(x_i,x') r(x',x_i) dx' \right\},
\label{Gr_Opt_Prob1}
\end{equation}
where,
\begin{equation}
\bar u^{-x_i}=\int_0^{T_f} g(\bar \gamma (z),z)(p(z)-r(z,x_i)) dz,
\end{equation}
and (with a slight abuse of notation) $r$ denotes the density of the measure $r(\cdot,x_i)$ with respect to the Lebesgue measure.

The optimization problem \eqref{Gr_Opt_Prob1}, assuming $\bar u^{-x_i}$ as given, can be reformulated as an optimal control problem using the state $x'$ as a virtual time.
\begin{proposition}
The optimization problem \eqref{Gr_Opt_Prob1} is equivalent to the optimal control problem:
\begin{equation}
\begin{aligned}
& \underset{u^{x_i}(t)}{\text{minimize}}
& &  \int_0^{T_f} L^{x_i} (u^{x_i}, \bar  u^{-x_i}+\chi_2^{x_i},t)dt\\
& \text{subject to}
& & \dot \chi _1^{x_i} = g(u^{x_i},t)r(t,x_i), \text{~~~~ } \chi_1(0)=0\\
& \text{~}
& & \dot \chi_2^{x_i}=0, \text{~~~~ } \chi_2^{x_i}(0):\text{free}\\
& \text{~}
& & \chi_1^{x_i}(T_f)=\chi_2^{x_i}(T_f),
\end{aligned}
\end{equation}
where 
\begin{equation}
L^{x_i} (u,v,t) = J \left (u,v, t\right) w(x_i,t) r(t,x_i) \nonumber
\end{equation}
\end{proposition}
\textit{Proof:} Observe that
$\int_0^{T_f} g(u^{x_i},t)r(t,x_i) dt= \chi_1^{x_i}(T_f)=\chi_2(T_f)=\chi_2(t)$.  \hfill $\square$

Necessary conditions can be derived using Pontryagin minimum principle (e.g. \cite{liberzon2011calculus}). It turns out that the problem has a special structure and the optimal control law is characterized by a pair of algebraic equations instead of a two point boundary value problem. The Hamiltonian is given by:
\begin{equation}
H^{x_i} = L^{x_i}(u^{x_i},\bar u^{-x_i}+\chi_2^{x_i},t) +p_1^{x_i} g(u^{x_i},t)r(t,x_i).
\end{equation}
The costate equations are given by:
\begin{align}
\dot p_1^{x_i}=0, ~~ \dot p_2^{x_i} =- \frac{\partial L^{x_i}}{\partial v}(u^{x_i},\bar u^{-x_i}+\chi_2^{x_i},t) 
\end{align}
(where $v$ is the second argument of $L^{x_i}$) and the boundary conditions by:
\begin{align}
p^{x_i}_2(0)=0,~~~  p_1^{x_i}(T_f)+p_2^{x_i}(T_f)=0. 
\end{align}

Let us assume that there is a unique minimizer $u^{x_i}=l(t,\chi_2,p_1,\bar u^{-x_i},x_i)$ of $H^{x_i}$ with respect to $u^{x_i}$. In order to characterize the optimal controller it remains to determine  the constants $p_1^{x_i}$ and $\chi_2^{x_i}$. A pair of algebraic equations will be derived.

Combining $\dot p_1^{x_i} =0$, $p_2^{x_i}(0)=0$ and $p_1^{x_i}(T_f)=-p_2^{x_i}(T_f)$ we get:
\begin{equation}
p_1^{x_i}=-p_2^{x_i}(T_f) = \int_0^{T_f} \frac{\partial L^{x_i} (l(t,\chi_2^{x_i},p_1^{x_i},\bar u^{-x_i},x_i), \chi_2^{x_i}+\bar u^{-x_i},t)}{\partial v} dt \label{alg_1}
\end{equation}
The right hand side of \eqref{alg_1} is a known function of $\chi_2^{x_i}$, $p_1^{x_i}$ and $\bar u^{-x_i}$.

The second algebraic equation is obtained combining $\dot \chi_2^{x_i}=0$, $\chi_1^{x_i}(0)=0$ and $\chi^{x_i}_1(T_f)=\chi^{x_i}_2(T_f)$:
\begin{equation}
\chi_2^{x_i} = \int_0^{T_f} g(l(t,\chi_2^{x_i},p_1^{x_i},\bar u^{-x_i},t,x_i) r(t,x_i) dt, \label{alg_2}
\end{equation} 
where the right hand side of \eqref{alg_2} is again a known function of $\chi_2^{x_i}$, $p_1^{x_i}$ and $\bar u^{-x_i}$.

\begin{proposition} 
Assume that $\bar \gamma(x)$ is an $r$-Kant-Nash equilibrium. Further assume that there is a unique minimizer $l$ of $H^{x_i}$ for  any $x_i\in X$. Then, there exist functions $\chi_2^{\cdot}:X\rightarrow \mathbb{R}$, $p_1^{\cdot}:X\rightarrow \mathbb{R}$ and $\bar u^{-\cdot}:X\rightarrow \mathbb{R}$ satisfying, \eqref{alg_1}, \eqref{alg_2} and:
\begin{equation}
\bar u^{-x_i}=\int_0^{T_f} g((l(t,\chi_2^{t},p_1^t,\bar u^{-t},x_i),t)(p(t)-r(t,x_i)) dt, \label{Int_eq}
\end{equation}
such that $\bar \gamma(x_i) = l(x_i,\chi_2,p_1,\bar u^{-x_i},x_i)$ for any $x_i\in X$.
\end{proposition}
\textit{Proof:} Immediate.  \hfill $\square$

Thus, an $r$-Kant-Nash equilibrium is characterized by a couple of algebraic equations and an integral equation.

\subsection{Equilibrium in a Quadratic Game}

Let us consider again the Fishing Game example assuming players with different efficiencies (for example a fisher is more experienced than another or he has a better boat). We assume that $\Theta$ is a singleton, $X=[0,1]$ and the players have a uniform distribution. The cost function for each player is given by:
\begin{equation}
J_i = u_i^2-(1-\bar u)\xi(x_i)u_i,
\end{equation}
where:
\begin{equation}
\bar{u} = \int_0^1 u(x)\xi (x) dx
\end{equation}
and $\xi(x)>0$ is the efficiency of a player with state $x$. 

We shall compute the $r$-Kant-Nash equilibrium assuming that $r(x',x)=0$ implies $w(x',x)=0$, that is if a player with state $x$ considers a player another player with state $x'$ to belong to his virtual group, he does not assign him a zero weight. 

In this example, the optimal control problems are LQ and thus the minimum principle necessary conditions are also sufficient.
The Hamiltonian is given by:
\begin{equation}
H^{x_i} = \left[u^2-(1-\bar u^{-{x_i}}-\chi^{x_i}_2)\xi(t) u \right]w(t,x_i)r(t,x_i)+p^{x_i}_1\xi(t)r(t,x_i)u
\end{equation}

The optimal control $u$ is given by:
\begin{equation}
u=l(t,\chi_2^{x_i},p_1^{x_i},u^{-{x_i}})=\frac{1}{2}(1-\bar u^{-{x_i}}-\chi_2^{x_i} -p_1^{x_i}/w(t,x_i))\xi(t)
\end{equation}

Equation \eqref{alg_2} is written as:
\begin{equation}
\chi_2^{x_i} = \frac{1}{2}\int_0^1 (1-\bar u^{-x_i}-\chi_2^{x_i} -p_1^{x_i}/w(t,x_i)) \xi^2 (t) r(t,x_i) dt
\end{equation}
or:
\begin{equation}
\chi_2^{x_i} = \frac{(1-\bar u^{-x_i})C_1^{x_i} -p_1C_2^{x_i}}{2+C_1^{x_i}}  \label{Alg_cond_Fish1}
\end{equation}
where:
\begin{equation}
C_1^{x_i}=\int_0^1  \xi^2 (t) r(t,x_i) dt \text{ ~ and ~ } C_2^{x_i}= \int_0^1  \xi^2 (t) r(t,x_i)/w(t,x_i) dt
\end{equation}

Equation \eqref{alg_1} is written as:
\begin{equation}
p_1^{x_i} = \frac{1}{2} \int_0^1  ((1-\bar u^{-x_i}-\chi_2^{x_i})w(t,x_i) -p_1)\xi^2(t) r(t,x_i) dt.
\end{equation}
Equivalently:
\begin{equation}
2p_1^{x_i} =  (1-\bar{u}^{-x_i})C_3^{x_i} -\chi_2^{x_i}C_3^{x_i}-p_1^{x_i}C_1^{x_i},
\label{Alg_cond_Fish2}
\end{equation}
where:
\begin{equation}
C_3^{x_i}=\int_0^1  \xi^2 (t) r(t,x_i) w(t,x_i)dt. 
\end{equation}

Solving \eqref{Alg_cond_Fish1}, \eqref{Alg_cond_Fish2} for $\chi_2^{x_i}$, $p_1^{x_i}$ we obtain:
\begin{align}
\chi_2^{x_i}  &= \frac{(C_1^{x_i})^2 +2C_1^{x_i}-C_2^{x_i}C_3^{x_i} }{(C_1^{x_i})^2 +2C_1^{x_i}-C_2^{x_i}C_3^{x_i}+4}(1-\bar{u}^{-x_i}), \\p_1^{x_i}&= \frac{2C_3^{x_i} }{(C_1^{x_i})^2 +2C_1^{x_i}-C_2^{x_i}C_3^{x_i}+4}(1-\bar{u}^{-x_i}).
\end{align}

In what follows, in order to simplify the computations we assume that $w(x,x')=1$. Under this assumption, it holds $C_1^{x_i}=C_2^{x_i}=C_3^{x_i}=C(x_i)$ and:

\begin{align}
\chi_2^{x_i}=p_1^{x_i}&= \frac{C(x_i) }{2C(x_i)+2}(1-\bar{u}^{-x_i}).
\end{align}
Furthermore,
\begin{align}
u^{x_i}(t) = \frac{1}{2}(1-\bar{u}^{-x_i})\frac{\xi(t)}{C(x_i)+1}.
\end{align}

Equation \eqref{Int_eq} becomes:
\begin{equation}
\bar u^{-x_i} = \int^1_0 \frac{1}{2}(1-\bar{u}^{-t})\frac{\xi^2(t)}{C(t)+1}(1-r(t,x_i)) dt, \label{integralLQ}
\end{equation}
which is a linear 	Fredholm integral equation of second kind.

\begin{example}
In this example we assume that $r(x,x')=\alpha$ (a uniform (sub)-distribution). Equation \eqref{integralLQ} implies that $\bar u^{-x_i}$ is independent of $x_i$. Thus, denoting by $\bar u^{-}$ this constant we obtain:
\begin{equation}
\bar u^{-} =(1-\bar{u}^{-}) \frac{1-\alpha}{2}\int^1_0 \frac{\xi^2(t)}{C(t)+1} dt.  
\end{equation}
Thus,
\begin{align}
u^{x_i}(x_i) = \frac{1}{2+(1-\alpha)\int^1_0 \frac{\xi^2(t)}{C(t)+1} dt}\frac{\xi(x_i)}{C(x_i)+1}.
\end{align}
Hence, the actions of the players scale down uniformly as $\alpha$ increases. \hfill $\square$

\end{example}

\begin{example}
In this example we assume that: 
\begin{equation}
r(x,x')=\begin{cases} \alpha &\mbox{if }  |x-x'|\leq 0.3 \text{~ and ~} x\leq0.9\\ 
0  &\mbox{otherwise } 
\end{cases}
\end{equation}
The solution of the integral equation \eqref{integralLQ} can be approximated using a linear system with a high order. The actions of the players, as well as the cost for the participants of the game are illustrated in figures \ref{EX_Actions},\ref{EX_Costs}. 
\begin{figure}
\includegraphics[width=\textwidth]{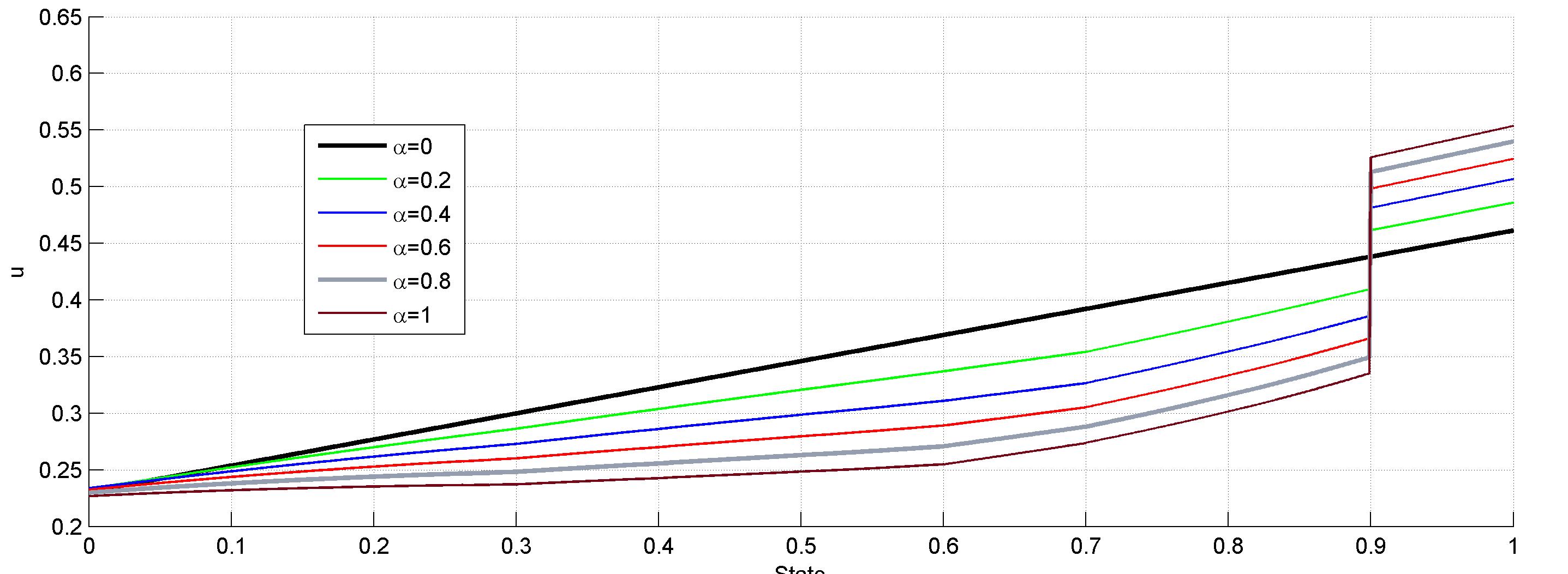}
\caption{The actions of the several players for different values of $\alpha$}
\label{EX_Actions}
\end{figure}
\begin{figure}
\includegraphics[width=\textwidth]{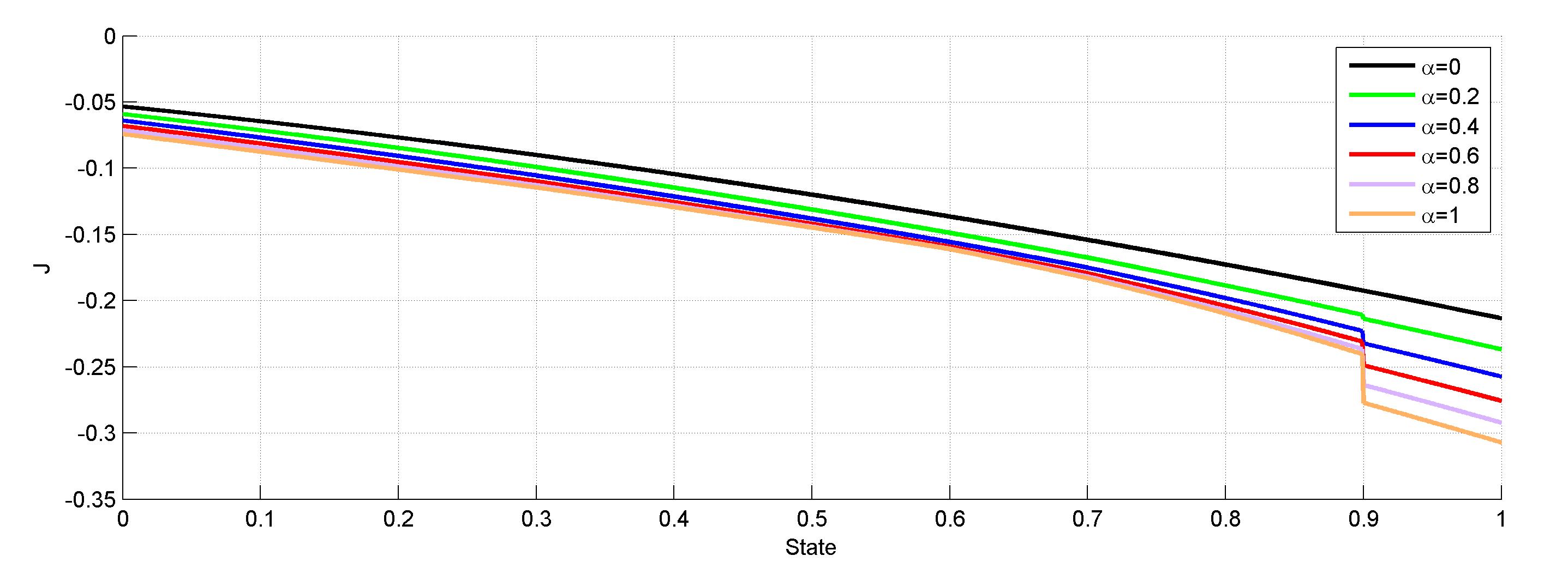}
\caption{The cost for the several players for different values of $\alpha$}
\label{EX_Costs}
\end{figure}

 \hfill $\square$

\end{example}

\section{An example of an $h,r$-Kant-Nash equilibrium}

We consider again the Fishing Game example involving players with different efficiencies who value differently their  time. Assume that $X=\{1,2\}\times\{1,2\}$ and that:
\begin{equation}
J_i = x_i^2u_i^2-(1-\bar u)x_i^1(x_i)u_i,
\end{equation}
where $x_i^1$ is the efficiency of Player $i$ and $x_i^2$ determines how much Player $i$ values her working time.
Further, assume that $p(1,1)=p_1=0.1$, $p(1,2)=p_2=0.2$, $p(2,1)=p_3=0.3$ and $p(2,2)=p_4=0.4$. 

In this section, we shall compare the $r$-Kant-Nash equilibrium with the $h,r-$Kant-Nash equilibrium, assuming that there is only a single type of social preferences $\theta$ and that $r(\cdot,x,\theta)$ is a uniform sub-probability measure with total mass $\alpha$.

\begin{figure}
\includegraphics[width=\textwidth]{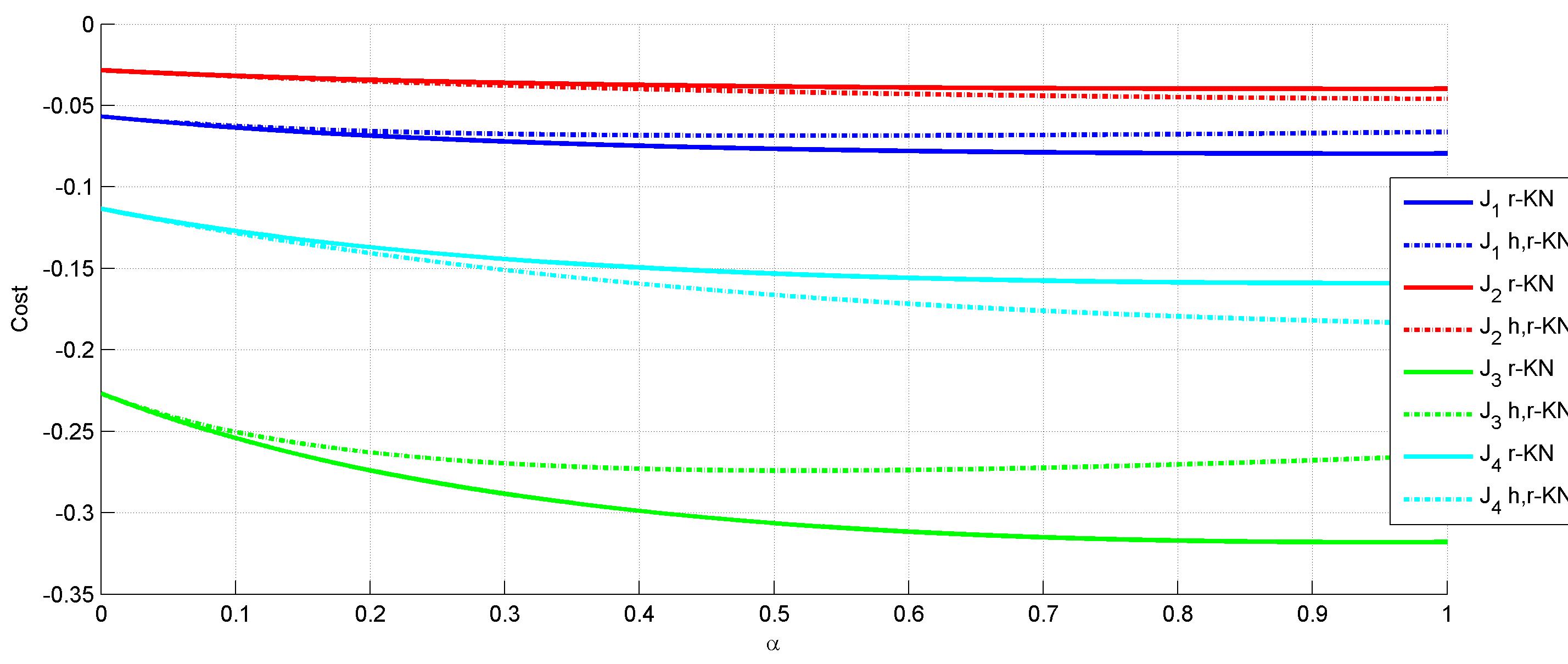}
\caption{The cost of the players of several types for the cases of $r$-Kant-Nash and $h,r$-Kant-Nash equilibria}
\label{h_r_fig1}
\end{figure}

\begin{figure}
\includegraphics[width=\textwidth]{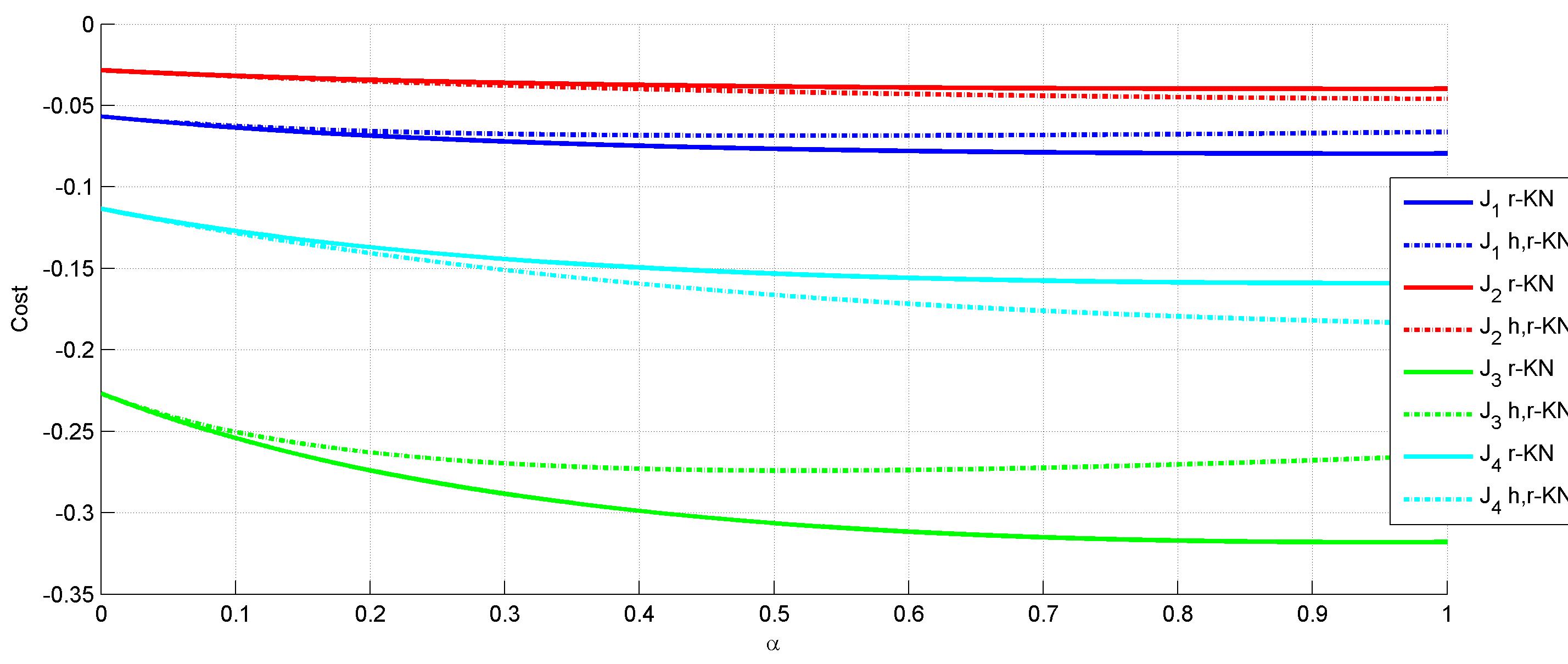}
\caption{The actions of the players of several types for the cases of $r$-Kant-Nash and $h,r$-Kant-Nash equilibria}
\label{h_r_fig2}
\end{figure}

\begin{example}[The $r$-Kant-Nash equilibrium] 
In order to derive the  equations characterizing a set of strategies $\gamma(1,1)=u_1^{KN}, \gamma(1,2)=u_2^{KN},\gamma(2,1)=u_3^{KN}, \gamma(2,2)=u_4^{KN}$ constituting an $r$-Kant-Nash equilibrium, consider the cost of the virtual group of player $i$:
\begin{align}
\tilde{J}_i &= p_1u_1^2+2p_2u_2^2+p_3u_3^2+2p_4u_4^2-\big[1-\alpha (p_1u_1+p_2u_2+2p_3u_3+2p_4u_4)-\nonumber\\\nonumber&~~~~~~~~-(1-\alpha)(p_1u_1^{KN}+p_2u_2^{KN}+2p_3u_3^{KN}+2p_4u_4^{KN})\big]\cdot \\\nonumber&~~~~~~~~~~~~~~~~~~~~~~\cdot(p_1u_1+p_2u_2+2p_3u_3+2p_4u_4), 
\end{align}
or in compact form:
\begin{equation}
\label{fish_rkn_cost}
\tilde{J}_i  =u^T(S+\alpha l l^T)u-(1-(1-\alpha)l^T{u}^{KN})l^Tu,
\end{equation}
where $S=\text{diag}(p_1,2p_2,p_3,2p_4)$, $l=[p_1,p_2,2p_3,2p_4]^T$, 
${u}^{KN}=[u_1^{KN},u_2^{KN}, u_3^{KN},u_4^{KN}]^T$ and ${u}=[u_1,u_2,u_3,u_4]^T$. A set of strategies $u^{KN}$ is an $r$-Kant-Nash equilibrium if the $u^{KN}$ minimizes \eqref{fish_rkn_cost} with respect to $u$. Equivalently, $u^{KN}$ is an $r$-Kant-Nash equilibrium if: \sloppy
\begin{equation}
(S+\alpha ll^T)u^{KN} = \frac{1}{2}(1-(1-\alpha)l^Tu^{KN})l
\end{equation}
After straightforward calculations we get:
\begin{equation}
u^{KN} = (2S+(1+\alpha)ll^T)^{-1}l
\end{equation}
The actions of the players and the costs are illustrated in Figures \ref{h_r_fig1} and \ref{h_r_fig2}. \hfill $\square$
\end{example}

\begin{example}[The $h,r$-Kant-Nash equilibrium]
Consider then the function $h:X\times X \rightarrow X$ with $h(({x'}^1,{x'}^2),(x^1,x^2))= ({x'}^1,x^2) $. That is, each player acts as is a potion $\alpha$ of the other players would follow her strategy taking into account the effectiveness of the actions of the others but not how they value their time\footnote{ For example a player of type $(2,1)$ i.e., an efficient player how has a small value for her  working time, may accept not to overfish because she doesn't want also the others to overfish taking into account the efficiency of the others. However, she may not be wiling to take into account that some of the other fishers value more their working time. That is, she may not be wiling to reduce her effort because some of the others are``lazy''.}. 

Let us then characterize an $h,r$-Kant Nash equilibrium $u^{KN}$. There are two kinds of virtual groups. The virtual group of a player with $x^2_i=1$ has cost:
\begin{align}
\tilde J_i =& p_1 (u_1^{t1})^2+p_2 (u_2^{t1})^2+p_3 (u_3^{t1})^2+p_4 (u_4^{t1})^2- [1-\alpha(p_1u_1^{t1}+p_2u_2^{t1}+2p_3u_3^{t1}+\nonumber\\&+2p_4u_4^{t1})-(1-\alpha)(p_1u_1^{KN}+p_2u_2^{KN}+2p_3u_3^{KN}+2p_4u_4^{KN})]\cdot\nonumber\\\nonumber &\cdot(p_1u_1^{t1}+p_2u_2^{t1}+2p_3u_3^{t1}+2p_4u_4^{t1}),
\end{align}
where $u_1^{t1} - u_4^{t1}$ are the actions $\gamma((1,1),x_i),\gamma((1,2),x_i),\gamma((2,1),x_i), \gamma((2,2),x_i)$ in the virtual group of a player having $x^2_i=1$.
In compact form:
\begin{equation}
\tilde{J}_i=(u^{t1})^T(S_1+\alpha ll^T)u^{t1}-(1-(1-\alpha)l^Tu^{KN})l^Tu^{t1},
\label{h_r_KN1}
\end{equation}
where  $S_1=\text{diag}(p_1,p_2,p_3,p_4)$, $l=[p_1,p_2,2p_3,2p_4]^T$,
${u}^{KN}=[u_1^{KN},u_2^{KN}, u_3^{KN},u_4^{KN}]^T$ and $u^{t1}=[u^{t1}_1,u^{t1}_2,u^{t1}_3,u^{t1}_4]^T$. Similarly, the virtual group of a player with $x^2_i=2$ has cost: \sloppy
\begin{equation}
\tilde{J}_i=(u^{t2})^T(S_2+\alpha ll^T)u^{t2}-(1-(1-\alpha)l^Tu^{KN})l^Tu^{t2},
\label{h_r_KN2}
\end{equation}
where  $S_2=2\text{diag}(p_1,p_2,p_3,p_4)$  and $u^{t2}=[u^{t2}_1,u^{t2}_2,u^{t2}_3,u^{t2}_4]^T$. 

The values for $u^{t1}$ and $u^{t2}$ which minimize \eqref{h_r_KN1} and \eqref{h_r_KN2} respectively satisfy the following equations:
\begin{align*}
2(S_1+\alpha l l^T)u^{t1}+(1-\alpha)ll^Tu^{KN}=l,\\
2(S_2+\alpha l l^T)u^{t2}+(1-\alpha)ll^Tu^{KN}=l,\\
\end{align*}
Coupled with the consistency conditions $u_1^{KN} = u^{t1}_1$, $u_3^{KN} = u^{t1}_3$ and $u_2^{KN} = u^{t2}_2$, $u_4^{KN} = u^{t2}_4$, the $h,r-$Kant-Nash equilibrium is characterized by:
\begin{equation}
\left[\begin{matrix}
  2(S_1+\alpha ll^T) & 0 &(1-\alpha)ll^T \\
  0& 2(S_2+\alpha ll^T)  &(1-\alpha)ll^T \\
  \text{diag}(1, 0, 1, 0)&\text{diag}(0, 1, 0, 1)& -I_4
 \end{matrix}\right] \left[\begin{matrix} u^{t1}\\u^{t2}\\u^{KN}  \end{matrix}\right] =  \left[\begin{matrix} l\\l\\0 \end{matrix}\right]
\end{equation}
The actions of the players and the costs are illustrated in Figures \ref{h_r_fig1} and \ref{h_r_fig2}. \hfill $\square$
\end{example}

\section{Conclusion and Future Directions}

The notions of $r-$Kant-Nash equilibrium and $h,r$-Kant-Nash equilibrium were introduced and compared with other notions. Necessary conditions, based on a reduction to a set of optimal control problems, can be derived for cases of games where the possible states admit an 1-dimensional representation. Some examples of quadratic games with a finite number of types was analyzed and $r$-Kant-Nash and $h,r-$Kant-Nash equilibria were computed using systems of linear equations.

A possible extension of this work is to study games with a finite number of players. In this case we may assume that the virtual group of each player is stochastic and that each player determines her action before she learns the realization of her virtual group. Another direction for future research is to extend the current model to Dynamic Games.

\bibliographystyle{IEEEtran}
\bibliography{refs1}

\begin{thebibliography}{10}
\bibitem{roemerwe}
------, ``How we (do and could) cooperate.''
\end{thebibliography}

\end{document}